\documentclass[twocolumn]{aastex62}

\usepackage{apjfonts}

\shorttitle{V1148 Sgr}
\shortauthors{Bond et al.}

\newcommand{\CaI}{\ion{Ca}{1}}
\newcommand{\CaII}{\ion{Ca}{2}}

\newcommand{\NII}{\ion{N}{2}}

\newcommand{\Gaia}{{\it Gaia}}

\newcommand{\Spitzer}{{\it Spitzer}}

\newcommand{\kms}{{\>\rm km\>s^{-1}}}
\newcommand{\masyr}{\rm mas\, yr^{-1}}

\begin{document}

\title{Nova Sagittarii 1943 (V1148 Sgr): A Luminous Red Nova?}

\author[0000-0003-1377-7145]{Howard E. Bond}
\affil{Department of Astronomy \& Astrophysics, Pennsylvania State University, University Park, PA 16802, USA}
\affil{Space Telescope Science Institute, 
3700 San Martin Dr.,
Baltimore, MD 21218, USA}

\author[0000-0003-3594-1823]{Jessica Mink}
\affil{Center for Astrophysics | Harvard \& Smithsonian, 60 Garden Street, Cambridge, MA 02138, USA}

\author{Alison Doane}
\affil{Center for Astrophysics | Harvard \& Smithsonian, 60 Garden Street, Cambridge, MA 02138, USA}
\affil{Deceased 2017 October~29}

\author{Sarah Lavallee}
\affil{Center for Astrophysics | Harvard \& Smithsonian, 60 Garden Street, Cambridge, MA 02138, USA}

\correspondingauthor{Howard E. Bond}
\email{heb11@psu.edu}

\begin{abstract}

Nova Sagittarii 1943 (V1148~Sgr) was an 8th-mag optical transient that was unusual in having a late-type spectrum during its outburst, in striking contrast to the normal high-excitation spectra seen in classical novae. Unfortunately, only an approximate position was given in the discovery announcement, hampering follow-up attempts to observe its remnant. We have identified the nova on two photographic plates in the Harvard archive, allowing us to determine a precise astrometric position. Apart from these two plates, obtained in 1943 and 1944, none of the photographs in the Harvard collection, from 1897 to 1950, show V1148~Sgr to limits as faint as $g\simeq18.3$. Modern deep images show a candidate remnant at $i\simeq19.2$, lying only $0\farcs26$ from the site of the nova. V1148~Sgr may have been a luminous red nova (LRN), only the sixth one known in the Milky Way. However, it lacks the near- and mid-infrared excesses, and millimeter-wave emission, seen in other LRNe, leaving its nature uncertain. We urge spectroscopy of the candidate remnant.

\null\vskip 0.2in

\end{abstract}




\section{The Unusual Optical Transient \\  Nova Sagittarii 1943 \label{sec:intro} }

The 8th-magnitude optical transient Nova Sagittarii 1943 (V1148~Sgr) was discovered on material in the
Harvard College Observatory's Astronomical Photographic Glass Plate Collection\footnote{For more information: \url{https://library.cfa.harvard.edu/plate-stacks}}
by Margaret Mayall. Unfortunately, her discovery was presented only briefly in a meeting abstract several years later \citep{Mayall1949}, and to our knowledge no further information was ever published. Nevertheless, V1148~Sgr has remained of considerable interest because of the unusual spectroscopic characteristics that she described. Mayall's discovery was made on objective-prism plates, obtained with the 10-inch Metcalf telescope at Bloemfontein, South Africa. She described the spectrum of V1148~Sgr on the discovery photograph, taken on 1943 August~19, as resembling that of a late K-type star, with strong absorption lines at \CaI\ 4226~\AA\ and \CaII\ H and~K, and possibly absorption bands of TiO\null. Five days later, the \CaI\ and \CaII\ absorption continued to be strong, but the spectrum now had a ``banded'' appearance, with wide Balmer emission features. The final spectrum, on 1943 August~29, showed ``wide bright bands.'' These findings are extremely unusual for a classical nova (CN), which develops a high-excitation emission-line spectrum, without prominent absorption features, soon after the outburst \citep[e.g.,][]{Williams1991,Walter2012}. In retrospect, it seems possible that the bright emission bands described by Mayall at the final epoch were actually a misinterpretation of a spectrum of a very cool object with strong molecular {\it absorption\/} bands, as further discussed below.\footnote{Mayall was very familiar with the normal spectral evolution of CNe, having announced her objective-prism discoveries of five other novae in her 1949 abstract, along with many others during her long career. Thus it is significant that she did not describe any of her three plates of V1148~Sgr as exhibiting a normal nova spectrum.}

Unfortunately, Mayall's three objective-prism plates are no longer available in the Harvard Plate Collection, so her spectroscopic findings cannot be investigated. Moreover, Mayall provided only approximate coordinates for V1148~Sgr, which lies in an extremely crowded field in the Galactic bulge [Galactic coordinates $(l,b)=(5\fdg2,-3\fdg0)$]. Thus, until now, its precise location has been unknown. However, Mayall reported that the nova had also been detected on three direct plates in the Harvard collection, raising the possibility that it could be recovered from that material.  

\section{Astrometry \label{sec:astrometry} }

Of the three direct plates mentioned by Mayall, two are available in the Harvard archive. One of them is plate SB4269, obtained on 1944 August~20 (about one year after the light-curve maximum) with the 60-inch Rockefeller Reflector at Bloemfontein, which is of excellent quality. It clearly shows the nova, albeit at a faint level. The transient is marked in ink on the glass side of the plate, presumably by Mayall herself. We digitized this plate with a UMAX Technologies scanner, as described by \citet{Mink2006}. Using the resulting {\tt .fits} image, we identified about a dozen nearby reference stars that are contained in the recent \Gaia\/ Early Data Release~3 (EDR3; \citealt{Gaia2016,Gaia2021}), and measured their $(x,y)$ positions using the {\tt imexamine} task in IRAF.\footnote{IRAF was distributed by the National Optical Astronomy Observatories, operated by AURA, Inc., under cooperative agreement with the National Science Foundation.} We corrected their EDR3 positions from the \Gaia\/ 2000 epoch\footnote{We obtained the \Gaia\/ astrometry from the Vizier service at \url{https://vizier.u-strasbg.fr/viz-bin/VizieR-3?-source=I/350/gaiaedr3}, which provides positions adjusted to the 2000 epoch using the measured proper motions.} to the epoch of the plate, using the proper motions given in the \Gaia\/ catalog, and employed them to place a precise astrometric frame on the image. The rms scatter of the registration was $0\farcs027$ and $0\farcs019$ in right ascension and declination, respectively. The derived position of V1148~Sgr (J2000 equinox, 1944.6 epoch) is:

\smallskip

\centerline{
18:09:07.801, $-$25:59:13.41 .
}

\smallbreak

This position differs by some $24''$ from the approximate one given by Mayall. We note that \citet{Duerbeck1987} had attempted to identify the remnant of V1148~Sgr, and gave coordinates for a 14th-mag star; this was the brightest object close to the Mayall position. He stated that it is ``very probably not the exnova.'' This object is detected on the SB4249 plate (and was one of the \Gaia\/ reference stars\footnote{Duerbeck's star is at J2000 equinox and epoch position 18:09:05.847, $-$25:59:08.08, which is $27\farcs52$ away from the location of V1148\,Sgr.} used in our astrometric solution); it is clearly distinct from the nova. 

\section{Photometry \label{sec:photometry} }

The Digital Access to a Sky Century at Harvard (DASCH) project \citep{Grindlay2009} is a long-term effort aimed at digitizing many of the plates in the Harvard collection, and performing photometry on the images. The DASCH Lightcurve Access website\footnote{\url{http://dasch.rc.fas.harvard.edu/lightcurve.php}} provides the photometry (calibrated to the Sloan $g$ band) and image cutouts. We searched for plates covering the location of V1148~Sgr, finding some 191 images. Of these only one plate, RB12347, obtained on 1943 August~24, shows the nova. Stellar images on this plate are elongated, but nevertheless the nova is conspicuous, at magnitude $g=9.55\pm0.34$, according to the photometry provided by the DASCH website. The star's position, magnitude, and observation date are consistent with the data presented by Mayall. This bright object coincides with the transient seen a year later, but considerably fainter, on SB4269, as described in the previous section. Thus it is clearly V1148~Sgr.


The 60-inch Rockefeller plates are not being scanned at this time by the DASCH project, so the DASCH website does not provide a magnitude determination for the 1944 observation recorded on SB4269. Instead, we performed aperture photometry on the nova and about a dozen neighboring stars, using the image scan described in the previous section. We then calibrated the data using $g$ magnitudes from the Pan-STARRS1 catalog\footnote{Obtained using the image-access tool at \url{https://ps1images.stsci.edu/cgi-bin/ps1cutouts}. Note that the Harvard plates were obtained using unfiltered blue-sensitive emulsions, so that the $g$ bandpass is a fairly close approximation to their sensitivity function.} \citep{Chambers2016} for the neighbor stars. This analysis yielded a magnitude of $g=16.72\pm0.35$ for V1148~Sgr, at the date 1944 August~20. 

There are 138 pre-eruption plates available from the DASCH website, covering the range 1897 August~4 to 1943 August~2. We examined the image cutouts for these plates, and V1148~Sgr was not detected on any of them. After the outburst there are 51 plates, from 1944 August~23 to 1950 July~17, which likewise do not show the transient. The limiting $g$ magnitudes vary widely for these plates, but generally range from $\sim$15.5 to $\sim$16.5 for most of them. However, from 1930 to 1939, some 74 plates of the site were obtained with the 24-inch Bruce Doublet, and these plates reach as faint as $\sim$18.0 to 18.2~mag for the best ones. The last two plates taken in 1943 before the detection near maximum on August~19 were obtained on 1943 May~15 and August~2. The limiting $g$ magnitudes on these two plates are $\sim$15.9 and 15.5, respectively. The DASCH database contains no plates between the one on 1943 August~24
showing the nova near maximum and the SB4269 plate obtained on 1944 August~20. Two more plates were obtained in 1944, both of them on August~23; neither one shows the nova; however, their limiting magnitudes are only $\sim$15.5 and 16.3. The best post-outburst plate, A26840, was obtained on 1949 June~8, and does not show the nova to a limiting magnitude of $\sim$18.3.

\begin{figure*}[ht]
\centering
\includegraphics[width=\textwidth]{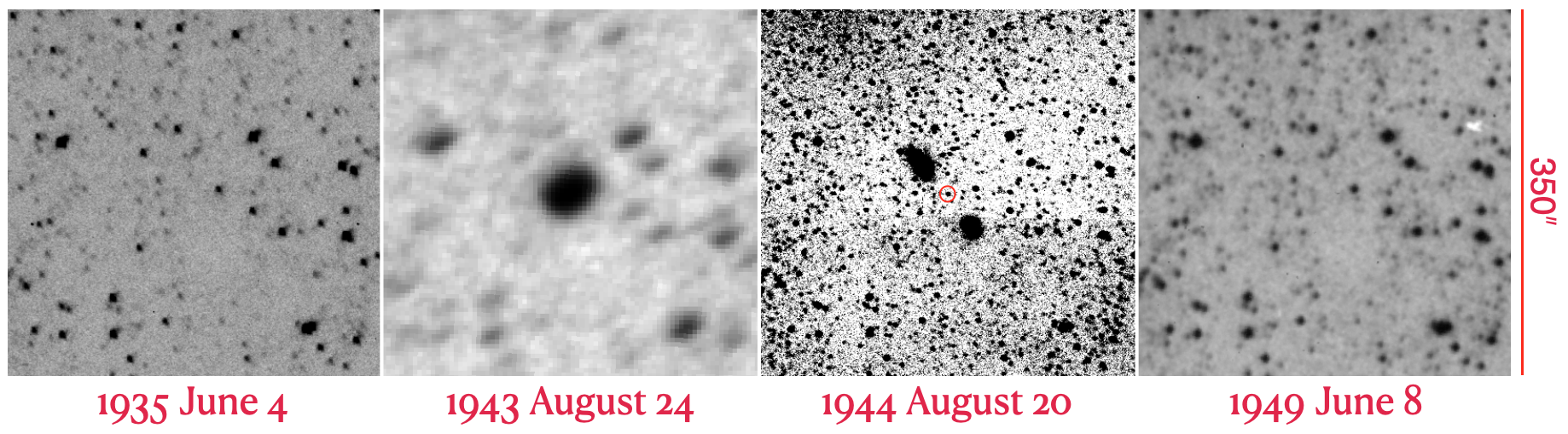}
\caption{
Image cutouts from digital scans of four plates from the Harvard College Observatory's Astronomical Photographic Glass Plate Collection, centered on the site of the transient V1148~Sgr. Each frame is $350''$ high; north at the top, east on the left. Outskirts of the globular cluster NGC\,6553 are visible at the northeast corners of the frames. {\it Left panel:} 1935 June~4. V1148~Sgr is not detected to a $g$ magnitude of $\sim$17.3. {\it Second panel:} 1943 August~24. The nova is near maximum brightness, at magnitude 9.55. {\it Third panel:} 1944 August~20. V1148~Sgr (between the two ink marks, and encircled in red) is detected at 16.7 mag. {\it Fourth panel:} post-outburst plate, taken on 1949 June~8. V1148~Sgr is fainter than $g\simeq18.3$~mag. 
\label{fig:mosaic}
}
\end{figure*}

Figure~\ref{fig:mosaic} shows four image cutouts from Harvard plates. The left panel shows the nova site on plate A17599 (24-inch Bruce Doublet, 1935 June~4). The limiting magnitude on this plate is about 17.3, based on reference to the Pan-STARRS1 sequence described above. There is no object detected at the nova position. 
The second frame in the figure 
is from the plate RB12347 described above (3-inch Ross Fecker telescope, 1943 August~24). Here the nova is conspicuous, near maximum light, in spite of the shallow limiting magnitude of $\sim$14.6 because of the trailed stellar images on this plate. 
The cutout in the third panel of Figure~\ref{fig:mosaic} shows the nova (between the two ink marks) on plate SB4269 (1944 August~20); this is the plate used for the astrometry and photometry described above. 
The fourth panel in the figure shows a cutout from the excellent plate A26840, obtained on 1949 June~8. As noted above, the limiting detection magnitude on this image is about 18.3, and the nova is not seen.

In summary, there are two Harvard direct plates showing V1148~Sgr, obtained on 1943 August~24 (mag 9.55) and 1944 August~20 (mag 16.72). These are in addition to the three objective-prism detections reported by \citet{Mayall1949} on 1943 August~19, 24, and~29, at a reported brightest magnitude of~$\sim$8. The rise to maximum occurred sometime between 1943 August~2 and 19. The deepest pre-outburst plates do not show a progenitor brighter than $\sim$18.0--18.2~mag. By mid-1949 the nova was fainter than $g$ magnitude~18.3.

\section{A Candidate Remnant \label{sec:remnant} }

\begin{figure*}[!]
\centering
\includegraphics[width=4in]{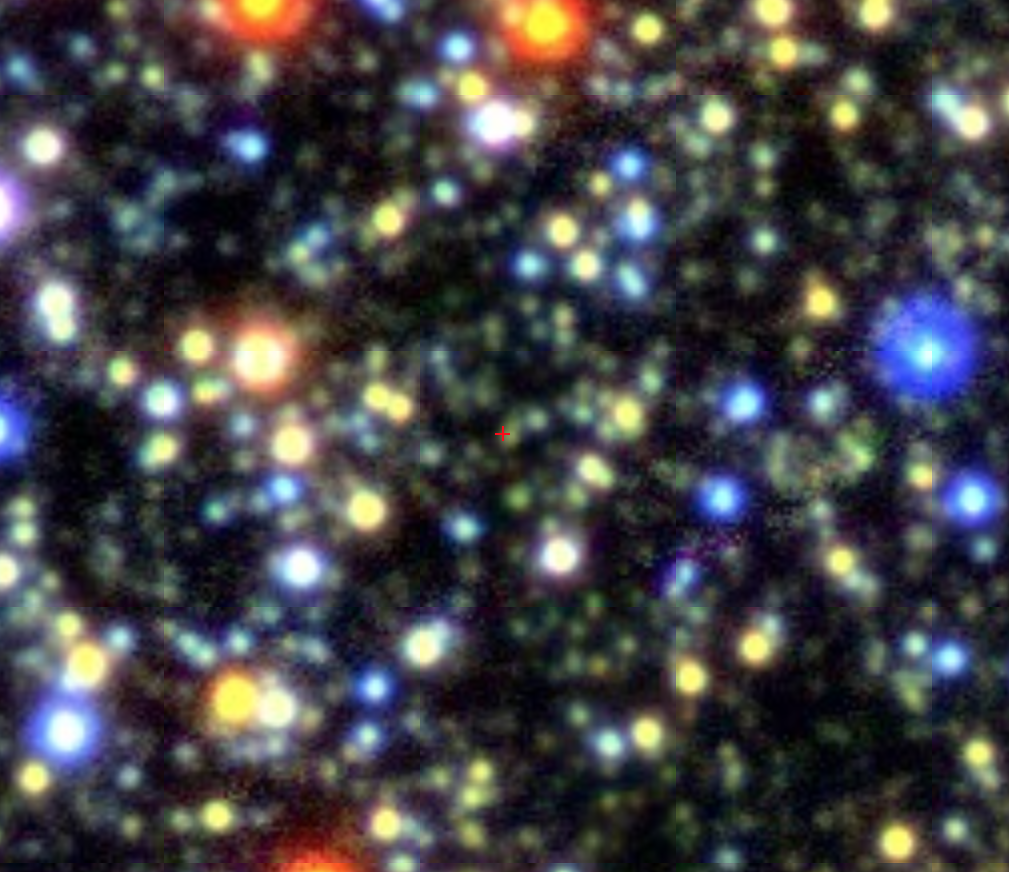}
\caption{
Color rendition of the field surrounding the site of V1148~Sgr, created using the
PanSTARRS-1 Image Access tool. Frame is $55''$ high, and has north at the top, east on the left. Red, green, and blue correspond to the $y$, $i$, and $g$ filters, respectively. A small red cross marks the astrometric position of V1148~Sgr. The immediately adjacent star to the southwest, with $i$ magnitude 19.18, is a candidate remnant, as discussed in the text. 
\label{fig:candidate}
}
\end{figure*}

Figure~\ref{fig:candidate} shows a modern deep color optical image of the site of V1148~Sgr, obtained from the Pan-STARRS1 website cited above. The red cross at the center of the frame marks the astrometric position of the nova, given in \S\ref{sec:astrometry}. Immediately southwest of the position is a faint star. This object is contained in the \Gaia\/ EDR3 catalog, with magnitudes of $G=19.81$ and $G_{\rm RP}=18.41$, lying at a J2000 position of

\smallskip

\centerline{
18 09 07.782, $-$25 59 13.47 ,
}

\smallbreak

\noindent corresponding to a separation of only $0\farcs26$ from the astrometric position of the nova. This star is thus a candidate remnant of V1148~Sgr. Unfortunately, \Gaia\/ EDR3 does not list a parallax or proper motion for the object. The star is also contained in the Pan-STARRS1 photometric catalog, with magnitudes of $(r,i,z,y) = (19.98, 19.18, 18.58, 18.29)$.

There is also a fainter star $0\farcs97$ north-northwest of, and partially blended with, the candidate in Figure~\ref{fig:candidate}. Neither the candidate nor the neighbor appear to have unusually red (or blue)  colors. As the figure shows, the Galactic-bulge field in which V1148~Sgr lies is extremely crowded, and it is entirely possible that these two stars are merely chance interlopers. Spectroscopic observations would provide further constraints.

Note that, because of the unknown proper motion of the nova, its current position could be slightly offset from the epoch 1944.6 location. For example, assuming a tangential velocity of $100\,\kms$, the displacement from 1944.6 to 2015 would be $\sim$$1\farcs5$ and $\sim$$0\farcs2$ for distances of 1000 and 8000~pc, respectively.

\section{Membership in Globular Cluster NGC\,6553?}

As noted by \citet{Mayall1949}, V1148~Sgr lies near the Galactic globular cluster NGC~6553,\footnote{Due to a typographical error, Mayall designated the cluster NGC\,6533.} whose outskirts can be seen at the northeast corner of the frames in Figure~\ref{fig:mosaic}. This raises the question whether the nova could be a member of the cluster. Only one, or possibly two, CNe are known to have occurred in the Galactic globular-cluster system \citep[e.g.,][and references therein]{Doyle2019}.

V1148~Sgr is located $300''$ from the center of NGC~6553. The tidal radius of the cluster, calculated from  information in \citet{Harris2010},\footnote{Online version of 2010 December, at \url{http://physwww.mcmaster.ca/~harris/mwgc.dat}} is $\sim$$460''$. Thus cluster membership is not ruled out. However, as Figure~\ref{fig:candidate} shows, the nova site is immersed in a rich Galactic-bulge field population. Color-magnitude and proper-motion diagrams of stars in \Gaia\/ EDR3 lying within $90''$ of V1148~Sgr show a nearly pure bulge population, with very few cluster members.\footnote{The color-magnitude diagram of the cluster, from the \Gaia\/ catalog, exhibits a clump of very red horizontal-branch stars, and the \Gaia\/ proper motions of the cluster stars are tightly concentrated. These make it easy to distinguish the members from the field.} 

A tighter constraint on cluster membership would be provided by the proper motion of the nova. If the faint star noted above is the remnant, then by combining its \Gaia\/ position with our astrometry from the 1944 plate, we calculate a proper motion of $(\mu_\alpha,\mu_\delta) = (-4.5, -1.1) \, \masyr$. This is in poor agreement with the mean proper motion of the cluster, for which we find $(\mu_\alpha,\mu_\delta) = (+0.4, -0.5) \, \masyr$ from data in the \Gaia\/ catalog.

In summary, the sparse available information casts doubt on, but does not absolutely rule out, a physical association of V1148~Sgr with the globular cluster NGC~6553. It is possible that a future data release from \Gaia\/ may provide a parallax, and a proper motion independent of the astrometry of the nova, for the candidate remnant. This could produce a more critical membership test---but only if evidence emerges that the candidate described in \S\ref{sec:remnant} is in fact the remnant.

\section{Was V1148~Sgr a Luminous Red Nova?}

Luminous red novae (LRNe) are a class of dust-forming astrophysical transients that has been recognized in recent years. CNe become increasingly blue as their eruptions proceed and their ejecta become optically thin, exposing a hot central source. LRNe erupt suddenly, like CNe, but they become extremely cool and red after their outbursts, due to an expanding photosphere and dust formation in the outflow. LRNe are generally considered to be the result of binaries undergoing common-envelope interactions and stellar mergers (see, for example, recent papers by \citealt{Ivanova2013}, \citealt{Pastorello2019}, and \citealt{Howitt2020}, and references therein); this was definitely the case for the Galactic LRN V1309~Sco, which was shown to have been a close binary with a rapidly decreasing orbital period before its eruption \citep{Tylenda2011}.

The known LRNe that have occurred in the Milky Way are V4332~Sgr \citep{Martini1999, Bond2018}, V838~Mon \citep{Sparks2008,Woodward2021}, OGLE-2002-BLG-360 \citep{Tylenda2013}, CK~Vul \citep[Nova Vul 1670;][]{KaminskiCKVul2021}, and V1309~Sco. In the Local Group, three LRNe have been recorded in M31: M31~RV \citep{Bond2011}, M31LRN 2015 \citep{MacLeod2017}, and AT~2019zhd \citep{Pastorello2021}. LRNe are often associated with old populations, including elliptical galaxies and the bulges of the Milky Way and other spiral galaxies; however, V838~Mon arose from a young open star cluster \citep{Afsar2007}, and several extragalactic LRNe appear to have originated from massive stars \citep[e.g.,][]{Smith2016,Blagorodnova2017, Blagorodnova2021}. (There is an extensive literature on LRNe, and we have cited only a few relevant and recent papers.)

In the case of V1148~Sgr, the late-type spectrum at the discovery epoch described by \citet{Mayall1949} may make it a candidate for another Galactic LRN,\footnote{A Galactic-bulge microlensing event is ruled out by the changing spectrum reported by Mayall.} similar to prototypes like V4332~Sgr and V838~Mon---if so, only the sixth one known in the Milky Way. Indeed, several authors have made the suggestion that V1148~Sgr was a LRN, including \citet{Bond2006}, \citet{Retter2006}, and \citet{Smith2016}. The apparent similarity of V1148~Sgr to V838~Mon and other red transients was noted by \citet{Kimeswenger2007}, who made an unsuccessful attempt to identify its remnant by searching the vicinity for H$\alpha$-emitting objects. However, our candidate, described in \S\ref{sec:remnant}, is considerably fainter than the objects he considered.

The fragmentary observations discussed in \S\ref{sec:photometry} indicate that V1148~Sgr declined some $\sim$10 magnitudes over the first year since its outburst, at blue wavelengths. This is not dramatically different from V838~Mon's drop of about 9~mag in $B$ in the first year after its eruption \citep[e.g., Figure~5 in][]{Sparks2008}.

Mayall's description of the nova's spectrum, at the final objective-prism observation, as having bright emission bands (see \S\ref{sec:intro}) remains a puzzle. A normal CN does rapidly develop a spectrum with broad, high-excitation emission lines (see the references cited in \S\ref{sec:intro}). However, Mayall did not characterize the spectrum as being that of a normal CN, indicating that it was unusual. It is conceivable that in fact the spectrum had, by late August 1943, become that of an extremely cool object, as is the case for LRNe. For example, \citet[][their Figure~1]{Barsukova2007} and \citet[][their Figure~2]{Munari2007} show late-time blue-region spectra of V838~Mon that are dominated by extremely strong molecular absorption. These can give the spectrum a superficial appearance of having strong emission bands, in the form of the continuum levels between the molecular bands. Unfortunately, since Mayall's spectrum plates are now lost, it is not possible to verify this interpretation. 

An argument against V1148~Sgr being a LRN comes from the fact that all of the known recent Galactic LRNe became conspicuous near- and mid-infrared sources following their outbursts: V4332~Sgr \citep{Banerjee2015}, V838~Mon \citep{Woodward2021}, OGLE-2002-BLG-360 \citep{Tylenda2013}, and V1309~Sco \citep{Tylenda2016}. The remnant of the much older event CK~Vul is not bright in the mid-infrared, but shows a far-infrared flux excess \citep{Evans2002}. 

\begin{figure*}[!]
\centering
\includegraphics[width=\textwidth]{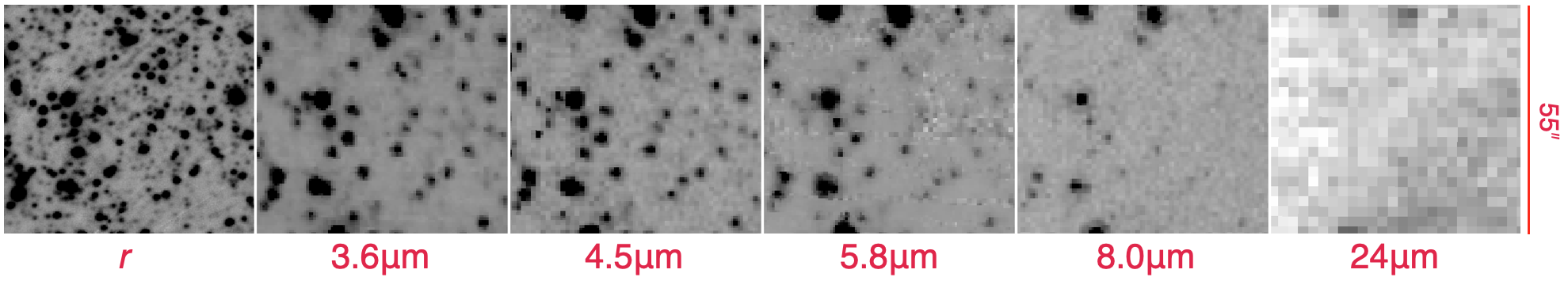}
\caption{
Images of the site of V1148~Sgr at the optical $r$ band (from PanSTARRS-1) and at five near- and mid-infrared wavelengths (from the {\it Spitzer Space Telesceope}). Frames are centered on the position of the transient, have the same $55''$ height as in Figure~\ref{fig:candidate}, and have north at the top, east on the left. No near- or mid-infrared counterpart of the nova is detected.
\label{fig:spitzermosaic}
}
\end{figure*}

To investigate whether there is a near- or mid-infrared source at the site of V1148~Sgr, we downloaded ``Super-Mosaic'' images obtained with the {\it Spitzer Space Telescope\/} from the NASA/IPAC Infrared Science Archive.\footnote{\url{https://irsa.ipac.caltech.edu/data/SPITZER/Enhanced/SEIP/overview.html}}  In Figure~\ref{fig:spitzermosaic} we show frames with the same $55''$ angular height as in Figure~\ref{fig:candidate}. The left-hand image is in the optical ground-based $r$ band, from the PanSTARRS-1 website, centered on the coordinates of the nova. The remaining panels show {\it Spitzer\/} frames from the Infrared Array Camera at 3.6, 4.5, 5.8, and $8.0\,\mu$m, and from the Multi-Band Imaging Photometer at $24\,\mu$m. These images demonstrate that there is no bright near- or mid-infrared source at the location of V1148~Sgr. The candidate remnant discussed in \S\ref{sec:remnant} is only marginally detected in the two shortest-wavelength \Spitzer\/ frames.

In addition, CK~Vul, along with V4332~Sgr, V838~Mon, and V1309~Sco, all lie at the centers of cool, compact molecular nebulae, which are prominent at millimeter wavelengths \citep{Kaminski2018,KaminskiCKVul2021,KaminskiV838Mon2021}.  \citet{Kaminski2022} attempted to detect CO emission from V1148~Sgr with the Atacama Pathfinder Experiment (APEX) 12-m submillimeter telescope, with unsuccessful results. Unfortunately, they pointed at the Duerbeck position discussed in \S\ref{sec:astrometry}. This would have placed the nova outside the APEX beam size, which ranged from $13\farcs2$ to $18\farcs7$. However, Kami{\'n}ski has informed us privately that his team has also observed V1148~Sgr with the Submillimeter Array\footnote{\url{https://lweb.cfa.harvard.edu/sma/}} (SMA), with a beam large enough to include the revised position reported here. No detection was made in the SMA data.


\section{Summary and Future Studies}

We have used Harvard plate material to determine a precise astrometric position for V1148~Sgr, an optical transient that appears to have had some spectroscopic features consistent with the handful of other members of the class of LRNe. However, this interpretation remains questionable because of the absence of a near- and mid-infrared or millimeter source at its location. Thus the nature of this transient remains unclear. 

We have identified a faint candidate remnant of the event for which a spectroscopic study would be useful. However, the observation would be challenging because of the source's faintness, as well as a neighboring star $\sim$$1''$ away (which itself may be another candidate counterpart).  

The LRN CK~Vul is surrounded by a faint ``hourglass'' nebula seen in narrow-band H$\alpha$ and [\NII] images \citep[][and references therein]{KaminskiCKVul2021}. Thus deep optical imaging of V1148~Sgr in these bandpasses would be of interest.

\acknowledgments

H.E.B. remembers with gratitude his introduction to the Harvard plate collection, nearly five decades ago, by Dr.~Martha L. Hazen, who became a cherished friend.

We thank Tomek Kami{\'n}ski for providing results from his SMA observations in advance of publication.

The DASCH project at Harvard is grateful for partial support from National Science Foundation grants AST-0407380, AST-0909073, and AST-1313370.

This research has made use of the VizieR catalogue access tool, CDS,
 Strasbourg, France (DOI : 10.26093/cds/vizier). The original description 
 of the VizieR service was published in 2000, A\&AS 143, 23.

This work has made use of data from the European Space Agency (ESA) mission
{\it Gaia\/} (\url{https://www.cosmos.esa.int/gaia}), processed by the {\it Gaia\/}
Data Processing and Analysis Consortium (DPAC,
\url{https://www.cosmos.esa.int/web/gaia/dpac/consortium}). Funding for the DPAC
has been provided by national institutions, in particular the institutions
participating in the {\it Gaia\/} Multilateral Agreement.

The Pan-STARRS1 Surveys (PS1) and the PS1 public science archive have been made possible through contributions by the Institute for Astronomy, the University of Hawaii, the Pan-STARRS Project Office, the Max-Planck Society and its participating institutes, the Max Planck Institute for Astronomy, Heidelberg and the Max Planck Institute for Extraterrestrial Physics, Garching, Johns Hopkins University, Durham University, the University of Edinburgh, the Queen's University Belfast, the Harvard-Smithsonian Center for Astrophysics, the Las Cumbres Observatory Global Telescope Network Incorporated, the National Central University of Taiwan, the Space Telescope Science Institute, the National Aeronautics and Space Administration under Grant No.\ NNX08AR22G issued through the Planetary Science Division of the NASA Science Mission Directorate, National Science Foundation Grant No.\ AST-1238877, the University of Maryland, Eotvos Lorand University (ELTE), Los Alamos National Laboratory, and the Gordon and Betty Moore Foundation.

This research has made use of the NASA/IPAC Infrared Science Archive, which is operated by the Jet Propulsion Laboratory, California Institute of Technology, under contract with the National Aeronautics and Space Administration.






\end{document}